\documentclass[
superscriptaddress,
twocolumn,
 amsmath,amssymb,
 aps,
prl,
a4paper,
dvipdfmx,
floatfix
]{revtex4-2}
\usepackage{graphicx,color}% Include figure files
\usepackage{dcolumn}% Align table columns on decimal point
\usepackage{bm}% bold math
\usepackage{siunitx}
\usepackage[version=4]{mhchem}
\newcommand{\Pdd}{\textit{X}[Pd(dmit)$_{2}$]$_{2}$ }

\everymath{\displaystyle}
\usepackage{here}
\begin{document}
\title{Power-Law Suppression of Phonon Thermal Transport by Magnetic Excitations in a Molecular Quantum Spin Liquid}
\author{S.~Fujiyama}
\email{fujiyama@riken.jp}
\affiliation{RIKEN, Pioneering Research Institute (PRI), Wako 351-0198, Japan}
\author{K.~Ueda}
%\affiliation{RIKEN, Condensed Molecular Materials Laboratory, Wako 351-0198, Japan}
\affiliation{National Institute of Technology, Anan College, Anan 774-0017, Japan}
\author{Y.~Otsuka}
\affiliation{RIKEN, Center for Computational Science (R-CCS), Kobe 650-0047, Japan}
\date{\today}
\begin{abstract}
We present large-scale \textit{ab initio} phonon calculations for the molecular
quantum spin liquid \textit{X}[Pd(dmit)$_2$]$_2$. An unusually low average phonon velocity ($\sim 700~\mathrm{m/s}$) and optical modes below $10~\mathrm{cm}^{-1}$ confine the Debye $T^{3}$ regime to $T \lesssim 2~\mathrm{K}$. As the transfer-integral anisotropy approaches the maximally frustrated regime ($t'/t\!\to\!1$), the lattice stiffens, ruling out lattice softening as the origin of the spin-liquid state. By quantifying the additional suppression of the thermal conductivity from experimental data, we observe a power-law behavior consistent with two-dimensional magnetic excitations with a nodal, approximately linear (Dirac-like) spectrum.
\end{abstract} 
%\pacs{75.25.-j, 75.70.Tj, 71.30.+}
%\keywords{Suggested keywords}%Use showkeys class option if keyword display desired
\maketitle
\newpage

%\section{Introduction}

Geometrical frustration in triangular-lattice antiferromagnets prevents classical spin ordering and gives rise to a quantum spin liquid (QSL) state, where macroscopic quantum fluctuations persist down to the lowest temperatures~\cite{Anderson1973,Balents2010,Savary2017,Zhou2017}. A key challenge in QSL research is to verify spin fractionalization and clarify whether the resulting spinons form a large Fermi surface~\cite{Anderson1973}. If these fractionalized spin excitations possess spectra analogous to those of a Fermi liquid, they should carry heat in the absence of conduction electrons. Because the electromagnetic response of spinons remains unresolved, most studies have relied on specific heat and thermal conductivity to probe a finite low‑temperature density of states (DOS).

\begin{figure}[hbt]
  \includegraphics*[width=0.9 \linewidth]{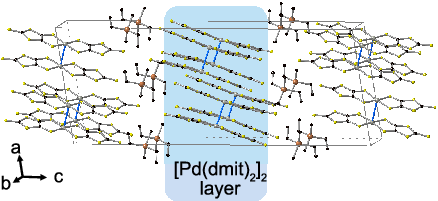}
  \caption{
Crystal structure of \textit{X}[Pd(dmit)$_2$]$_2$ (\textit{X} = Et$_2$Me$_2$Sb). The blue-shaded region highlights the two-dimensional magnetic layer composed of [Pd(dmit)$_2$]$_2$ dimers hosting $S = 1/2$ spins. These magnetic layers are separated by  closed-shell cation layers, giving rise to a quasi-two-dimensional structure.
}
  \label{fig:structure}
  \end{figure}

Molecular solids are known to realize QSL phenomena~\cite{Kanoda2011,Zhou2017,Powell2011}. Among them, EtMe$_{3}$Sb[Pd(dmit)$_{2}$]$_{2}$ (dmit = 1,3-dithiole-2-thione-4,5-dithiolate) consists of [Pd(dmit)$_{2}$]$_{2}^{-}$ dimers hosting $S = 1/2$ spins that form a triangular network. This two-dimensional magnetic layer is separated by closed-shell cation layers, as shown in Fig.~\ref{fig:structure}. The absence of a gap in the spin excitation spectrum has been widely reported~\cite{Itou2008,Itou2010,Kato2014,Fujiyama2018,Pustogow2018,Watanabe2012,Fujiyama2019,Oshima2024,Ueda2024,Ido2022}. Both the specific heat ($C$) and thermal conductivity ($\kappa$) exhibit finite values of $C/T$ and $\kappa/T$ in the low-temperature limit, suggesting the existence of a finite spinon density of states and possibly a large spinon Fermi surface~\cite{MYamashita2010,SYamashita2011}. 
However, the reported thermal conductivities differ significantly among experiments~\cite{Bourgeois-Hope2019,Ni2019}. Even crystals obtained from the same sample source exhibit serious discrepancies in $\kappa/T$, and the origin of these differences remains an open question. Although variations in cooling rate or microcracks affecting the mean free path have been proposed~\cite{Yamashita2019}, no corresponding changes in resistivity or $^{13}$C NMR spectra have been observed~\cite{Kato2022}. The unresolved nature of the spinons' electromagnetic response makes it difficult to reconcile the conflicting experimental observations.

Phonons in correlated electronic quantum materials have seldom been systematically investigated. Recent developments now allow \textit{ab initio} calculations of phonon dispersions and their contributions to thermal properties directly from crystal structures~\cite{Broido2009,Tadano2014}. Although fractionalized spin excitations cannot be evaluated within density-functional theory (DFT), phonon contributions can be computed to provide a quantitative lower bound for the measured thermal properties, thereby enabling indirect separation of the magnetic component.

Molecular solids such as \Pdd and BEDT-TTF-based salts provide several candidates for QSLs.
In these systems, the triangular lattice is generally anisotropic, and the ratio of transfer integrals $t'/t$ characterizes the deviation from an equilateral geometry.
Notably, the QSL phase can survive even when $t'/t$ deviates moderately from unity~\cite{Ueda2024}, indicating that spin correlations are strongly damped. Possible origins of this damping include partial delocalization of molecular orbitals near a metal--insulator transition and the intrinsic softness of the molecular lattice, which allows dynamical motion of localized spin sites~\cite{Manna2018}.
Phonons are therefore expected to play an essential role. However, the large unit-cell volume ($\sim1700~\text{\AA}^3$) and the presence of more than 100 atoms per cell render first-principles calculations extremely demanding, requiring up to $10^{4}$ self-consistent field calculations to evaluate $\kappa(T)$.

In this work, we report large-scale first-principles calculations of phonon dynamics and thermal properties for \Pdd (\textit{X} = \ce{Me4P} (AF), \ce{Me4As} (AF), \ce{EtMe3Sb} (QSL), and \ce{Et2Me2Sb} (CO)) across a wide range of triangular lattice anisotropy ($0.62 \leq t'/t \leq 1.005$). We discovered an intrinsic lattice softness characterized by remarkably low phonon group velocities ($\langle v \rangle \approx 700\text{ m/s}$), which is roughly $1/10$ of typical inorganic compounds, and the presence of very low-energy optical modes below $\omega < 10\ \text{cm}^{-1}$. This non-Debye behavior limits the validity of the $T^3$ specific heat regime to $T \lesssim 2\text{ K}$. Crucially, we observe increased $v$ as $t'/t \to 1$, revealing a lattice hardening towards the QSL regime. This result rules out lattice softening as a possible mechanism for the realization of the QSL state. 
By comparing our calculations with experiments, we quantify an additional suppression of $\kappa$ beyond phonon--phonon and boundary scattering, and find a power-law behavior consistent with two-dimensional Dirac-like magnetic excitations.

%\section{Calculation method}
The computational workflow begins with the geometry relaxation of
\textit{X}\ce{[Pd(dmit)2]2}.
The atomic structures were fully relaxed using the \textsc{Quantum ESPRESSO}
package~\cite{Giannozzi2009,Kato2012}.
For \textit{X} = \ce{EtMe3Sb}, where the two ethyl groups are orientationally disordered
with 50\% occupancy, we adopted the cation-ordered \textit{X} = \ce{EtMe3As} structure
measured below 230~K as the initial configuration~\cite{Kato2006}.
For \textit{X} = \ce{Et2Me2Sb}, which undergoes a first-order charge-ordering transition
at $T_\mathrm{CO}=73$~K, the relaxation was performed using the structure
observed above $T_\mathrm{CO}$.
Among the \textit{X}[Pd(dmit)$_2$]$_2$ family, this compound has the transfer-integral
ratio $t'/t$ closest to unity and is therefore regarded as representative of
the QSL state~\cite{Tsumuraya2013,Misawa2020}.
We employed the PBEsol exchange--correlation functional together with the
SSSP precision pseudopotentials~\cite{Prandini2018}.
Phonon dispersion, lattice specific heat, and thermal conductivity were
calculated using the \textsc{ALAMODE} package~\cite{Tadano2014}.
Further computational details are provided in Ref.~\cite{Supplement}.

\begin{figure}[hbt]
\includegraphics*[width=0.9\linewidth]{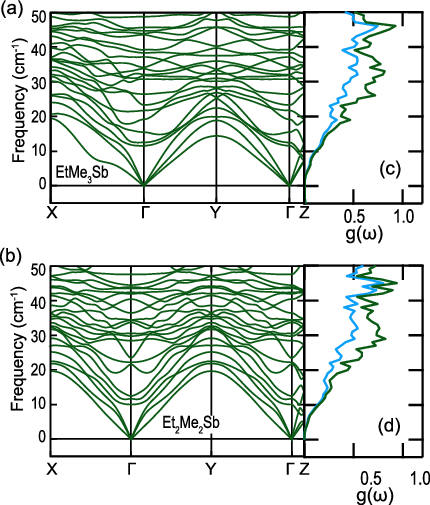}
\caption{
Phonon dispersions for \textit{X} = EtMe$_3$Sb (a) and Et$_2$Me$_2$Sb (b), calculated without assuming a charge-order transition.
The high-symmetry directions are $X=a^{*}-b^{*}$ and $Y=2a^{*}$ for (a), and $X(Y)=-a^{*}\pm b^{*}$ for (b), defined using the reciprocal vectors of the conventional cell.
(c),(d) Phonon densities of states (DOS) for the two salts, with the total DOS in green and the \ce{[Pd(dmit)2]2$^{-}$} anion contribution in blue.
Low-energy modes below 10 cm$^{-1}$ arise predominantly from vibrations within the anion layers.
}
\label{fig:dispersion}
\end{figure}

%\section{Results}
We plot in Fig.~\ref{fig:dispersion} the calculated phonon dispersions, $\omega(\mathbf{k})$, and the phonon density of states (DOS), $g(\omega)$ (see Ref.~\cite{Supplement} for \ce{Me4P} and \ce{Me4As}).
The group velocities of the acoustic phonons along the $\Gamma$–X(Y,Z) directions were evaluated from the slopes of the linear dispersions, $\omega = v_{X,(Y,Z)}k_{X,(Y,Z)}$, and are summarized in Table~\ref{tab:velocity}, where $k_{X,(Y,Z)}$ denotes the wave vector along $\Gamma$–X(Y,Z).
From the relation $g(\omega)/\omega^2 = V/(2\pi^2\langle v \rangle^3)$, we also estimated the weighted average velocity $\langle v \rangle$, where $V$ is the unit-cell volume.
The obtained velocities, $\langle v \rangle\approx 700$~m/s, are about one tenth of that of silicon ($v\approx 8$~km/s).
The lowest optical phonon modes also appear at energies as small as $\omega\lesssim 10$~cm$^{-1}$.
These very low-energy phonons reflect the intrinsic softness of molecular solids, producing an enhanced $g(\omega)$ at low energies and narrowing the frequency range where the Debye approximation ($\omega\propto k$) holds.

\begin{table}%[H] add [H] placement to break table across pages
\caption{Group velocities of \textit{X}\ce{[Pd(dmit)2]2}
\label{velocity}}
\begin{ruledtabular}
\begin{tabular}{cccccc}
\textit{X} & $t'/t$ & $\langle v \rangle \mathrm{(m/s)}$ & $v_X\mathrm{(m/s)}$ & $v_Y\mathrm{(m/s)}$ & $v_Z\mathrm{(m/s)}$ \\
\hline
\ce{Me4P} & 0.620 & 640 & 470 & 470 & 750 \\
\ce{Me4As} & 0.696 & 700 & 650 & 650 & 900 \\
\ce{EtMe3Sb} & 0.907 & 680 & (780) & (830) & (1600) \\
\ce{Et2Me2Sb} & 1.005 & 760 & 790 & 790 & 1110 
% Lines of table here ending with \\
\end{tabular}
\end{ruledtabular}
\label{tab:velocity}
\end{table}

We also plot the projected phonon density of states for the \ce{[Pd(dmit)2]2-} anions in Fig.~\ref{fig:dispersion}(c) and (d).
The low-energy phonons are dominated by vibrations within the \ce{[Pd(dmit)2]2} layers, suggesting that thermal transport is effectively confined to the two-dimensional magnetic planes.

The $v$'s clearly depend on the cation, \textit{X}.
The in-plane velocity $v_X$ increases as $t'/t \to 1$, revealing lattice hardening toward the most frustrated regime.
Although molecular solids are widely considered to possess soft lattices that may lead to positional instability of the localized spins, our results show that the lattice hardens as the system approaches the QSL regime.
This implies that the QSL behavior observed in \textit{X} = \ce{EtMe3Sb} originates primarily from geometrical frustration rather than lattice softness.

\begin{figure}[hbt]
  \includegraphics*[width=0.9 \linewidth]{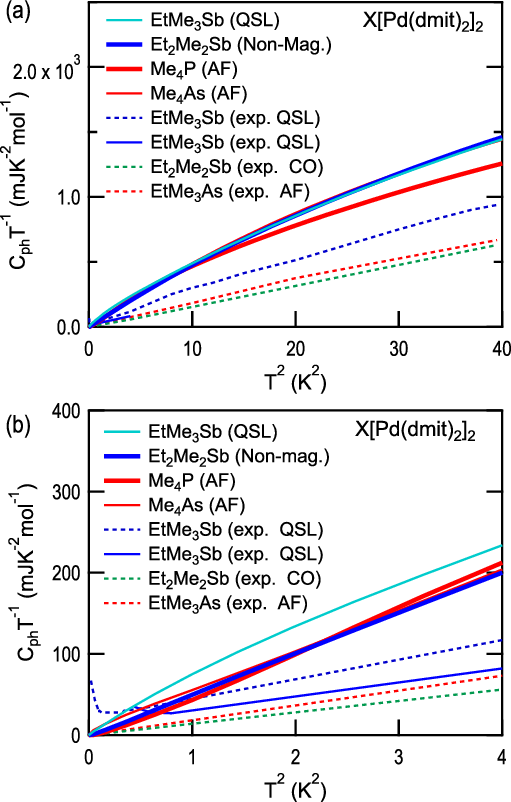}
  \caption{
Calculated $C_{\mathrm{ph}}/T$ as a function of $T^2$ for (a) $0 < T^2 < 40$ K$^2$ and (b) $0 < T^2 < 4$ K$^2$.
Experimental data for \textit{X} = EtMe$_3$As (AF), EtMe$_3$Sb (QSL), and Et$_2$Me$_2$Sb (CO, below the structural phase transition) are replotted from Reference \onlinecite{Nomoto2022}.
The calculated values are about twice as large as the experimental data, indicating that the phonon contribution alone overestimates the total specific heat, though the origin of this discrepancy remains to be understood.
  }

  \label{fig:Cph}
  \end{figure}

We plot in Fig.~\ref{fig:Cph} the calculated lattice specific heat divided by temperature, $C_{\mathrm{ph}}/T$, as a function of $T^2$ for (a) $0<T^2<40$~K$^2$ and (b) $0<T^2<4$~K$^2$.
Experimental data for \textit{X} = \ce{EtMe3As} (AF), \ce{EtMe3Sb} (QSL), and \ce{Et2Me2Sb} (CO, below the structural transition) are replotted from Ref.~\onlinecite{Nomoto2022}.
The calculated $C_{\mathrm{ph}}/T$ reproduces the experimental trend, while its absolute magnitude is larger by a factor of order two at the lowest temperatures.
Such a difference is reasonable given uncertainties in first-principles treatments of low-energy phonons and in relaxation-type specific-heat measurements on small single crystals, where full thermal equilibration between the sample and the thermometer can be difficult to achieve.
Importantly, this level of discrepancy does not affect the qualitative conclusions regarding the non-Debye phonon behavior.

The $C_{\text{ph}}/T$ exhibits pronounced nonlinearity with respect to $T^2$. Despite being purely the phonon contribution, a component with a power law smaller than $T^3$ becomes visible at $T \lesssim 2$ K ($T^2 \lesssim 4$ K$^{2}$). This suggests that the validity of the Debye approximation for lattice specific heat in this system is questionable at $T \gtrsim 2$ K, and the conventional analysis---extrapolating $C/T$ to $T=0\text{ K}$ to determine the electronic specific heat---is not applicable. This non-Debye behavior is directly linked to the lowest optical phonon branch falling significantly, to frequencies as low as $ \omega \lesssim 10\text{ cm}^{-1}$.

A comparison emphasizing the low-temperature regime ($0 < T^2 < 4\text{ K}^2$), however, shows that the experimental $C/T$ for $X = \text{EtMe}_3\text{Sb}$ still surpasses our calculated $C_{\text{ph}}/T$. Thus, our results do not fully exclude the possibility of a finite residual $C/T$ term as $T \to 0\text{ K}$.

Thermal conductivity ($\kappa/T$) provides a sensitive probe, complementary to the specific heat, for detecting fractionalized spins in QSLs.
An early report~\cite{MYamashita2010} found a finite $\kappa/T \approx 0.2~\mathrm{mW\,K^{-2}\,cm^{-1}}$ as $T \to 0$, interpreted as evidence for a large spinon Fermi surface, consistent with a finite residual $C/T$~\cite{SYamashita2011}.
Subsequent measurements, however, showed that $\kappa/T \to 0$ toward the lowest temperatures, revealing substantial discrepancies among different experiments~\cite{Bourgeois-Hope2019,Ni2019,Nomoto2022}. The lattice thermal conductivity is primarily governed by the effective phonon lifetime.
Within a Matthiessen-type framework, the total scattering rate can be written as
\begin{equation}
\frac{1}{\tau_{\mathrm{tot}}(T)}
=
\frac{1}{\tau_{\mathrm{ph\text{-}ph}}(T)}
+
\frac{2v}{L}
+
\frac{1}{\tau_{\mathrm{sp\text{-}ph}}(T)},
\label{eq:Matthiessen_full}
\end{equation}
where the three terms represent phonon--phonon scattering, boundary scattering with characteristic domain size $L$, and additional scattering arising from spin--phonon coupling, respectively.

\begin{figure}[t]
  \centering
  \includegraphics[width=0.95\linewidth]{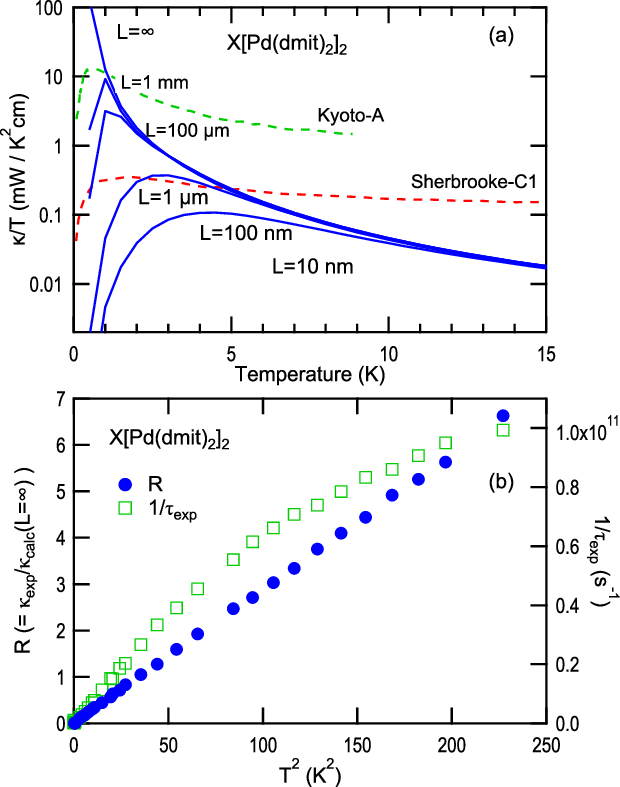}
  \caption{
(a) Calculated thermal conductivity $\kappa/T$ for infinite and finite domain sizes $L$, compared with experimental data from Ref.~\onlinecite{Bourgeois-Hope2019}.
The data reported in Refs.~\cite{Ni2019,Nomoto2022} nearly overlap with the Sherbrooke-C1 dataset.
A broad maximum around $T \approx 2$~K observed experimentally is reproduced for a characteristic domain size of $L \approx 100~\mu$m.
(b) The ratio $R=\kappa_{\mathrm{exp}}/\kappa_{\mathrm{calc}}(L=\infty)$ (left axis), together with the experimentally inferred phonon scattering rate $1/\tau_{\mathrm{exp}}$ (right axis), where $\tau_{\mathrm{exp}}$ is defined in Eq.~(\ref{eq:tauexp_def}).
The quadratic power-law behavior is most clearly observed in $R(T)$.
  }
  \label{fig:kOverT}
\end{figure}

Figure~\ref{fig:kOverT}(a) compares the calculated $\kappa/T$ for $L=\infty$ and finite $L$ with experimental data, considering only the first two terms in Eq.~(\ref{eq:Matthiessen_full}).
A broad maximum around $T \approx 2~\mathrm{K}$ is reproduced for $L \approx 100~\mu\mathrm{m}$, indicating that the peak temperature is primarily governed by macroscopic boundary effects.
The absolute magnitude of the experimental $\kappa/T$, however, is substantially smaller than the calculated phonon contribution.

To characterize the additional suppression of thermal transport beyond the phonon-only calculation,
we introduce the dimensionless ratio
\begin{equation}
R(T) \equiv \frac{\kappa_{\mathrm{exp}}(T)}{\kappa_{\mathrm{calc}}(T;L=\infty)},
\label{eq:R_def}
\end{equation}
which quantifies the reduction of the experimental thermal conductivity relative to the ideal phonon limit
without boundary scattering.
This definition removes the contribution from boundary scattering by construction and isolates
the suppression channel not captured by the phonon--phonon scattering term in
Eq.~(\ref{eq:Matthiessen_full}).

We define an effective experimental lifetime as
\begin{equation}
\tau_{\mathrm{exp}}(T) = \frac{\kappa_{\mathrm{exp}}(T)}{C_{\mathrm{ph}}(T)\,v^{2}},
\label{eq:tauexp_def}
\end{equation}
where $C_{\mathrm{ph}}(T)$ is the calculated phonon specific heat and $v$ is a characteristic sound velocity
($v \simeq 800~\mathrm{m\,s^{-1}}$).
The corresponding scattering rate $1/\tau_{\mathrm{exp}}$ provides a model-independent measure of the total
scattering strength inferred directly from experiment.

Figure~\ref{fig:kOverT}(b) shows $R(T)$ together with the experimentally inferred scattering rate
$1/\tau_{\mathrm{exp}}$.
While $R(T)$ selectively probes the additional suppression beyond the phonon-only calculation,
$1/\tau_{\mathrm{exp}}$ reflects the total scattering probability without assuming its microscopic origin.

In QSLs, the ground state is expected to possess strong singlet correlations, leading to a rapid reduction of thermally excited magnetic degrees of freedom at low temperatures.
If such magnetic excitations act as scatterers of phonons, the additional scattering rate $1/\tau_{\mathrm{sp\text{-}ph}}$ is expected to decrease upon cooling.
Accordingly, $R(T)$ directly reflects this additional suppression channel.
Although $\sqrt{R(T)}\,v$ is not identical to a mean free path, similar temperature dependences between $R(T)$ and $1/\tau_{\mathrm{exp}}$ are expected when spin--phonon scattering dominates.

Notably, $R(T)$ exhibits an approximately quadratic power-law behavior at low temperatures.
When attributed to spin--phonon scattering governed by the density of low-energy magnetic excitations,
this $T^{2}$ dependence is consistent with a nodal, approximately linear (Dirac-like) dispersion
in a two-dimensional QSL~\cite{Ido2022,Nomura2021,Oshima2024}.
This interpretation is further supported by the nuclear spin--lattice relaxation rate $1/T_{1}$,
which also reflects gapless magnetic excitations with nodes~\cite{Itou2011}.
It is also noteworthy that antiferromagnetically ordered salts exhibit substantially larger $\kappa$
than the QSL compound~\cite{Nomoto2022}.
This contrast is naturally understood if spin--phonon scattering is strongly suppressed in the ordered phase,
owing to the limited phase space of low-energy magnetic excitations,
in sharp contrast to the gapless excitation continuum characteristic of the QSL state.

%\section{Summary}
In summary, we have performed large-scale first-principles calculations to evaluate the phonon dispersion and lattice contributions to thermal transport in the molecular QSL $X[\mathrm{Pd(dmit)}_2]_2$.
Our results reveal pronounced lattice softness, characterized by an average phonon group velocity $\langle v\rangle \approx 700~\mathrm{m/s}$, an order of magnitude smaller than that in typical inorganic crystals, and the presence of optical modes below $10~\mathrm{cm}^{-1}$ ($\sim 1.2$~meV), which restrict the validity of the Debye regime to $T \lesssim 2~\mathrm{K}$.
We further find that the lattice stiffens as $t'/t \rightarrow 1$, ruling out lattice softening as the primary origin of the QSL state.
By quantifying the additional suppression of $\kappa$ beyond the phonon baseline, we extract its characteristic power-law temperature dependence, which provides strong evidence for low-energy magnetic excitations with a nodal, approximately linear (Dirac-like) dispersion.
These results demonstrate that first-principles evaluation of phonon transport is feasible even for molecular crystals with more than one hundred atoms per unit cell, and establish a quantitative route to connect lattice dynamics with emergent quantum magnetism.
The present framework is broadly applicable to correlated molecular solids, where lattice softness, magnetic frustration, and electronic correlations are intricately intertwined.

\begin{acknowledgments}
We are grateful to H.~Matsuura, M.~Ogata, H.~Seo, and T.~Tsumuraya for fruitful discussions. This work was supported by Grants-in-Aid for Scientific Research (24K06949, 25H02158, 24K08373, 	24K06894) from JSPS. This work used computational resources of the HOKUSAI supercomputer at RIKEN (RB230004, RB230063) and the Fugaku supercomputer provided by the RIKEN Center for Computational Science through the HPCI System Research project (hp240391).
\end{acknowledgments}
%References
%\begin{thebibliography}{apsrev4-1}
%\bibliographystyle{apsrev4-2}
%\bibliographystyle{apsrev-nourl}
%\end{document}

%\bibliographystyle{apsrev-nourl}
%\end{thebibliography}
\end{document}